\documentstyle[12pt,epsfig,a4]{article}
\begin{document}

\title{ Optimisation of on--line principal component analysis\\ }
\author{E Schl\"osser$^{\dag}$, D Saad$^{\ddag}$, and M Biehl$^{\dag}$
\\ $^{\dag}$ Institut f\"ur Theoretische Physik \\ Universit\"at
W\"urzburg, Am Hubland, \\ D-97074~W\"urzburg \\ $^{\ddag}$ The Neural
Computing Research Group \\ Aston University, Aston Triangle \\
Birmingham B4 7ET, U.K.}  \today

\maketitle

\begin{abstract}
Various techniques, used to optimise on-line principal component
analysis, are investigated by methods of statistical mechanics. These
include local and global optimisation of node-dependent learning-rates
which are shown to be very efficient in speeding up the learning
process. They are investigated further for gaining insight into the
learning rates' time-dependence, which is then employed for devising
simple practical methods to improve training performance.  Simulations
demonstrate the benefit gained from using the new methods.
\end{abstract}
\medskip

\underline{1.Introduction}

The investigation of unsupervised on-line learning
algorithms~\cite{Hertz,Bishop} by means of statistical mechanics has
been shown to be a useful tool for gaining insight on the training
dynamics~\cite{Bi94}.  In contrast to batch algorithms whereby all
available examples are considered simultaneously for calculating a
single student parameters update, on-line updates are carried out
after the presentation of each single data point (for an overview on
current on-line methods in neural networks see~\cite{Sa98}).  This
update is proportional to a learning rate $\eta$ that has to be
smaller than a critical value to make learning possible~\cite{WaAl93}.
Successful learning is only possible if the learning rate is
relatively small which, at the same time, means that many update steps
are needed. Therefore, a relatively large rate is needed at the
beginning and a smaller one later on; perfect learning is only
possible if $\eta \rightarrow 0$ at late stages of the learning
process.  For practical problems there is only empirical knowledge of
how the learning rate has to evolve \cite{Bishop}.  The use of variational
techniques \cite{KiCa92,SaRa97} enables one to calculate the optimal
learning rate evolution $\eta$ theoretically; however, these
calculations require information about the task and the input
distribution which is usually unavailable. Nevertheless, insight
gained from the analysis about the optimal learning rate time-dependence
may be used to improve training in practical scenarios.

There are mainly two learning rate optimisation paradigms which we
will discuss here: Local optimisation maximises the cost function loss
at every time step while global optimisation seeks the maximisation of
the cost function loss within a predetermined time window. Note that
towards the end of the time window the two methods coincide and that a
sufficiently long time window should be considered for the system to
converge. \newline

\underline{2. General Framework}

The algorithm examined here is an on-line-algorithm for principle
component analysis based on Sanger's rule \cite{Sanger}.  It was
already discussed in detail for constant learning rates $\eta_i$
\cite{BiSch98}. We consider here $N$-dimensional data vectors
$\underline{\xi}$ taken independently from a Gaussian
data-distribution with M relevant orthonormal directions $\{\underline
{B_i}\}_{i=1,...,M}$ ($M\ll N$, $ \underline{B}_{i}^\top
\underline{B}_{j} \, = \delta_{ij}$). The correlation matrix
$\underline{\underline{C}}=\left\langle \underline{\xi}
\underline{\xi}^\top \right\rangle$ of this distribution has the form
\begin{equation}
  \underline{\underline{C}} \, = \, \underline{\underline{I}}+
    \sum_{i=1}^{M}(b_{i}^{2}
    +2b_{i})\underline{B}_{i}\underline{B}_{i}^\top \  ,
\end{equation}
were $\{b_i\}_{i=1}^M$ are some positive parameters representing the
specific task and $\underline{\underline{I}}$ is the identity matrix.

In the on-line-scenario a single vector $\underline{\xi}^{\mu}$ is
presented every time step and a set of student vectors
$\underline{J}_l \in I\!\!R^N \quad (l=1,2,\ldots,M)$ is updated
according to
\begin{equation}
  \underline{J}_{l}(\mu)= \underline{J}_{l}(\mu-1)+\frac{\eta_{l}}{N}
  \, x_{l}^{\mu} \left(\underline{\xi}^{\mu}-\sum_{k=1}^{l}
  x_{k}^{\mu}\underline{J}_{k}(\mu-1)\right) \ ,
\end{equation}
with the student projections $ x_{l}^\mu = \underline{J}_l^\top
\underline{\xi}^\mu $. The student vectors are normalized explicitly 
after each time step.

In the limit $N \rightarrow \infty$ the evolution of the system can be
described by a set of coupled differential equations in `time' $\alpha
= \mu/N$ for the quantities $ R_{kl} (\mu) = \underline{J}_{k}^\top
(\mu) \underline{B}_{l} $ and $Q_{kl} (\mu)=\underline{J}_k^\top
(\mu)\underline{J}_l (\mu) $ which describe the overlaps of the
student vectors with the unknown principal components and the mutual
overlap:
\begin{eqnarray} \label{dgln}
 \frac{dR_{lj}}{d\alpha} & = & \eta_l \, \left\langle\,
   x_l y_j \,\right\rangle - (\eta_l+\eta_l^2/2) \, \left\langle\,
   x_l^2 \,\right\rangle R_{lj} \\ \hspace*{-25mm} & & - \eta_l \sum_{k=1}^{l-1} \,
   \left\langle\, x_l x_k \,\right\rangle (R_{kj}-Q_{lk}R_{lj}) \ \
   (k,l = 1,2,\ldots, M) \nonumber \\
   \frac{dQ_{lm}}{d\alpha} & = & (\eta_l+\eta_m)\, \left\langle\, x_l
   x_m \,\right\rangle -((\eta_l+\eta_l^2/2)\, \left\langle\, x_l^2
   \,\right\rangle + (\eta_m + \eta_m^2/2)\, \left\langle\, x_m^2
   \,\right\rangle ) Q_{lm} \nonumber \\ \hspace*{-25mm} & & - \eta_l \sum_{k=1}^{l-1}
   \, \left\langle\, x_l x_k \,\right\rangle (Q_{km} - Q_{kl}Q_{lm})
   -\eta_m \sum_{k=1}^{m-1} \, \left\langle\, x_m x_k \,\right\rangle
   (Q_{lk}-Q_{km}Q_{lm}) \ (l\neq m) \nonumber \ .
\end{eqnarray}
The averages over the quantities $x_k=\underline{J}_k^\top
\underline{\xi}$ and $y_j = \underline{B}_j^\top \underline{\xi}$ can
be performed analytically yielding
\begin{eqnarray} \label{cova}
   \, \left\langle\, x_k y_j \,\right\rangle = (1 + b_j)^2 R_{kj}, \quad \,
   \left\langle\, y_i y_j \,\right\rangle = (1+b_j)^2\delta_{ij},
   \mbox{~~and~~} \\
   \, \left\langle\, x_k x_l \,\right\rangle = Q_{kl} + \sum_{i}^{M}
   (b_i^2 + 2 b_i) R_{ki} R_{li} \nonumber \ .
\end{eqnarray}
An investigation of this learning scenario with constant learning
rates showed that the entire process depends crucially on the learning
rate.  Learning rates have to be slightly different for each student
vector to break symmetries which emerge between them during training
and to avoid time-consuming plateaus \cite{BiSch98}.  The values have
to be chosen between large learning rates which are suboptimal
asymptotically and small learning rates that result in prohibitively
slow learning at the transient.

To improve learning performance and speed it is necessary to choose 
time-dependent learning rates. As it was already shown that different
learning rates for different nodes are important \cite{BiSch98} we focussed on
finding appropriate solutions for a node-dependent learning rates
$\eta_l(\alpha)$. \newline

\underline{3. Locally Optimised Learning Rate}

One way to calculate an optimised learning rate is to maximise the
cost function loss in every time-step (local optimisation), i.e.,
obtaining $\eta_i(\alpha)$ from a minimisation of
$d\epsilon/d\alpha$ \cite{KiCa92}:
\begin{equation}
   \frac{\partial}{\partial \eta_l} \frac{d \epsilon}{d \alpha}=0 \ ,
\end{equation}
Choosing the cost function
\begin{equation}
\label{eq:costfunction}
   \epsilon=1-\frac{1}{M}\sum^{M}_{l=1}R_{ll}^{2} \ .
\end{equation}
This function is a measure of the learning success on a scale 
between 1 and 0, representing poor and optimal performance respectively, and
may be used to derive the locally optimal learning rate of the form
\begin{equation}
   \eta_l(\alpha)=-1+\frac{(1+b_l)^2}{(1+A_{ll})}-\frac{\sum_{k=1}^{l-1}
   (Q_{lk}+A_{lk})(R_{kl}-Q_{kl}R_{ll})}{(1+A_{ll})R_{ll}} \ ,
\end{equation}
with
\begin{equation}
   A_{kl}=\sum_{i=1}^{M}(b_i^2+2b_i)R_{ik}R_{il} \nonumber \ .
\end{equation}

\begin{figure}[t]
\centerline{
\epsfig{figure=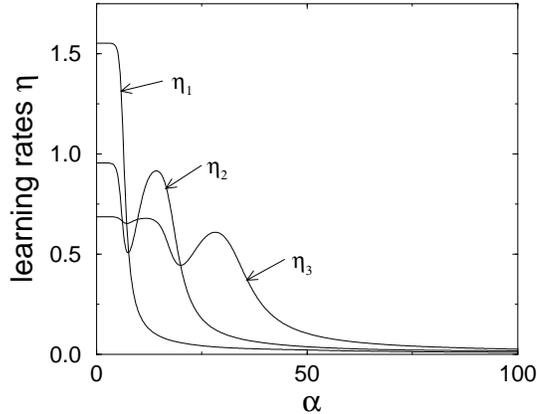,width=7.0cm}}
\caption{Time-dependence of the learning rates of the first three students
calculated with local optimisation.}
\label{first}
\end{figure}
\begin{figure}
\centerline{
\epsfig{figure=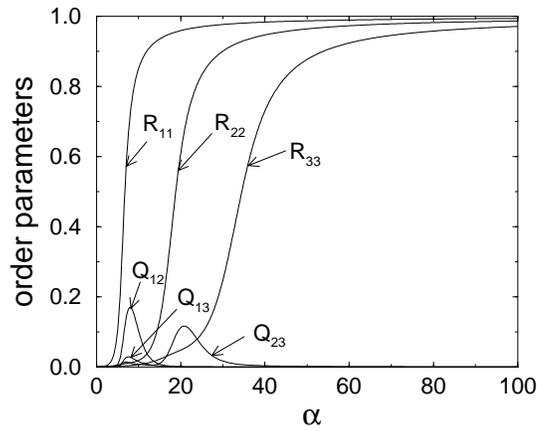,width=7.0cm}}
\caption{Overlap of the first three students with their principal components
and their mutual overlap, learning with the the locally optimised learning
rates shown above.}
\label{second}
\end{figure}
\begin{figure}
\centerline{
\epsfig{figure=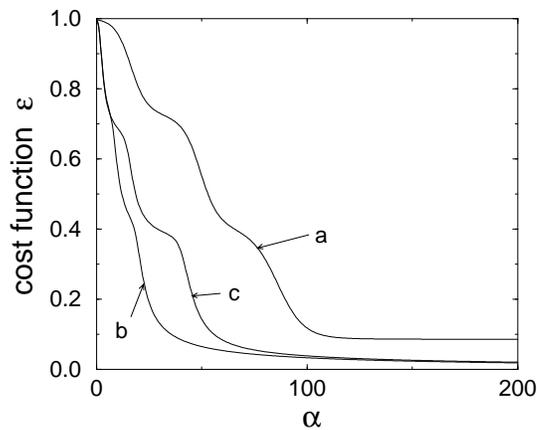,width=7.0cm}}
\caption{Cost functions for the detection of the first three
components in on-line PCA: the graph shows the learning process with
constant learning rates (a), with locally optimised (b) and with
globally optimised (c) learning rates.}
\label{third}
\end{figure}

This learning rate depends on the data structure and the order
parameters of the problem. By choosing these optimal learning rates,
the principal components are learnt very fast and high performance can
be achieved. In the following we choose a data distrbution (1) with
$b_1=0.6$, $b_2=0.4$ and $b_3=0.3$. 
Figure~\ref{first} shows the evolution of the learning
rates for the first three student vectors.  They all begin with a constant
value which depends on the data structure and have a decaying phase
later on where the learning rate decays roughly as $1/\alpha$. In
addition, the learning rates show a `dip' at the point where another
student vector learns the current, most-dominant, principal component
direction. This behaviour is explained by Figure~\ref{second},
showing the overlaps $R$ of the students with the principal components
(upper curves) and their mutual overlap $Q$ (lower curves):  The
principal components are learned one after another; all students try
to learn the largest p.c. first, which results in a significant
overlap $Q$ with the first student. The orthogonalisation realised by
the algorithm pushes the other students away from that direction to
specialise on other directions related to the less dominant p.c.  Once
the direction of the p.c. has been identified, the related learning
rate, of the specialised student vector, starts decaying.  At the same
time there is a significant overlap with the other student vectors
that started learning the same direction; consequently, their learning
rates are suppressed so as to prevent them from specialising on this
p.c.  any further and to facilitate the change in direction.

Figure~\ref{third} shows the evolution of the cost
function(\ref{eq:costfunction}). Curve (a) represents a learning
scenario with reasonably chosen constant learning rates
($\eta_1=0.1$, $\eta_2=0.108$ and $\eta_3=0.09$) 
balancing between training speed and asymptotic performance.  The
hierarchical structure of the learning process can be noticed here as
the three students learn the different principal components one after
the other.  The same learning process but with locally optimised
learning rates is shown in curve (b). The principal components are
learned very fast, resulting in very good asymptotic performance.
The locally optimised learning rate clearly provides improved
performance with respect to every constant rate. However, it depends
on knowledge that is not available in practical situations and can
therefore only provide insight into the optimal evolution of
$\eta_l$. \newline

\underline{4. Globally Optimised Learning Rate}

As the learning process may comprise different phases, for which local
optimisation may result in sub-optimal global performance, we will
also consider here a different approach based on global
optimisation~\cite{SaRa97} of the learning rate. This has been shown
to outperform local optimisation over a predetermined time
window. This method maximises the cost function loss over a fixed time
window:
\begin{equation}
\Delta \epsilon=\int^{\alpha_1}_{\alpha_0} d\alpha \
\left(\frac{d\epsilon}{d\alpha} -\sum_i \lambda_i \
(\mbox{constraints}) \right)
\end{equation}
where the constraints are the equations of motion (\ref{dgln}) which
have to be satisfied at every point in time and $\lambda_i$ are the
related Lagrange multipliers.  The time-window $\alpha_1 \!-\!
\alpha_0$ has to be chosen beforehand.  Applying a variational
approach with respect to the order parameters and their time
derivatives leads to a set of differential equations for the Lagrange
multipliers, from which the globally optimised learning rates $\eta_i$
can be derived.  Clearly, like in any other method, a minimal
time-window is required for the learning to converge.

Globally optimal parametrisation was shown to be much more efficient
in the case of plateaus in the learning process where local
optimisation leads to indefinite trapping~\cite{SaRa97}. However, one
has to keep in mind that global optimisation only looks at the total
loss $\Delta \epsilon$, so that intermediate values of $\epsilon$ can
be much worse than those obtained via local optimisation.  In the case
of on-line-PCA it turns out that that after the minimal time needed
for the algorithm to converge, the learning performance of locally and
globally optimised learning are similar. Figure~\ref{third} shows the
evolution of the learning process in both cases.

One can notice that the principal components are found later than with
local optimisation. This can be explained in the following way: If one
component is found more accurately, the orthogonalisation process can
push the other students much more efficiently out of that direction,
so that learning the next p.c. becomes easier. Local optimisation
does not rely on future gains and therefore chooses to carry on with
the specialisation of student vectors, providing better intermediate
performance.  The features of the globally optimised learning rates
are similar to those obtained via local optimisation.

Global optimisation would have been useful in the case of plateaus;
these emerge in the case of on-line-PCA only for a single learning
rate $\eta_i(\alpha)=\eta(\alpha)$ \cite{BiSch98}.  Therefore, in most
cases, global optimisation will not have any advantage over local
optimisation.

Note that instead of calculating a globally optimised learning rate
that leaves the learning rule itself unchanged, one can also calculate a
globally optimised learning rule \cite{RaSa97}. In our case this can
only be calculated numerically and does not provide additional information.
\newline

\underline{5. Discussion}

The use of locally optimised learning rates shows a significant
improvement in the learning performance over fixed learning rates, but
it depends on quantities that are not available in practical
applications of on-line-PCA. Nevertheless, insight gained from the
theoretical study may be useful for improving performance in practical
cases. From the analysis it is clear that the learning rates have to
be constant first and should decay like $1/\alpha$ at later times,
after specialisation took place. This point, where the learning rate
schedule should be changed has to be set through observables
accessible in practical scenarios; typically, one could use constant
learning rates until the asymptotic regime is reached, identified
through students stationarity. Our analysis provides a refined
criterion which leads to much faster learning: In figure~\ref{first}
and \ref{second} on notices that the decaying phase for a certain
student starts where the overlap to other students (usually to one in
particular) starts growing significantly. At this point the first
student has already learned the current most-dominant p.c.  and
becomes almost stationary while other students (one in particular),
which show significant correlation with the first student, should
start moving to other directions, being pushed away by the
orthogonalisation process.  The first student has learned enough to
stabilise and to begin its `fine tuning' which corresponds to the
stage of a decaying learning rate.  The mutual overlaps are the only
order parameters accessible in real applications; here they provide a
practical criterion for the starting point of the learning rate decay
phase, a criterion which was obtained directly from the analysis.
\newline

\underline{6. Simulation}
\begin{figure}[t]
\centerline{
\epsfig{figure=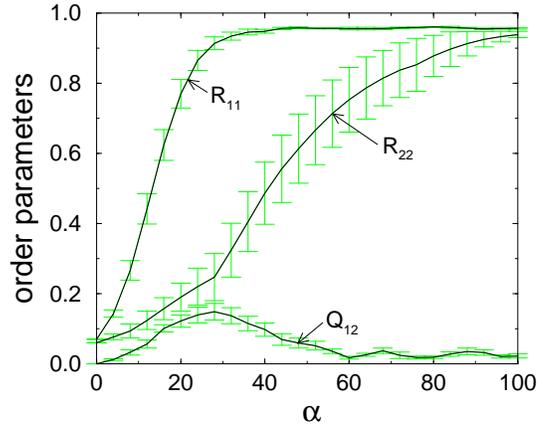,width=7.0cm}}
\caption{Simulation of on-line-PCA with constant learning rates:
Overlaps of the first two students with their PC's $R_{ll}$ and their
mutual overlap $Q$.}
\label{fifth}
\end{figure}
\begin{figure}
\centerline{
\epsfig{figure=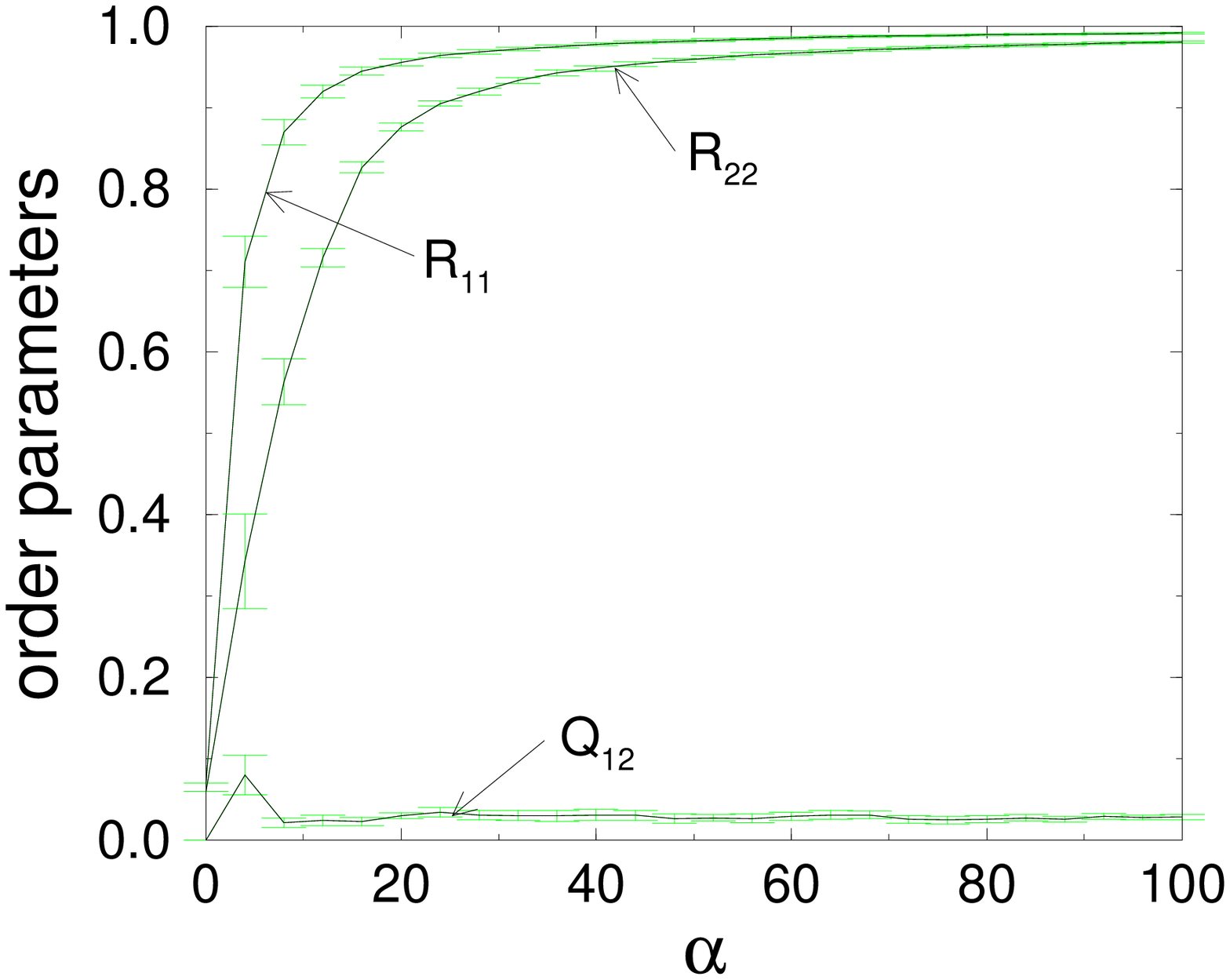,width=7.0cm}}
\caption{Simulation of on-line PCA like in figure~\ref{fifth}. The
learning rates are now chosen as time-dependent according to the
suggested rule. A comparison with figure 4 demonstrates clearly 
the efficiency of this method.}
\label{sixth}
\end{figure}

Simulations of an on-line principle component analysis were made to
test the usefulness of the criterion explained above. Fig~\ref{fifth}
displays a scenario with constant learning rates ($\eta_1=0.1$ and
$\eta_2=0.09$), learning the same data distribution as before.
The graph shows the overlaps of the first two students with the
corresponding principal components $R_{ll}$ and their mutual overlap
$Q$ as means and variances of ten runs.
The asymptotic regime is reached at the end of the time scale;
at this point one would normally commence the decay of the learning
rates. In comparison, we applied the rule suggested above, based on
monitoring the overlaps between student vectors, to the same task as
shown in figure \ref{sixth}. As soon as the overlap between two
students starts growing significantly the decay of the first student
commences; the decay for the next student commences according to
similar criteria. This corresponds directly to the observations of the
optimised learning process. We should point out that the starting
value of the learning rates can be chosen higher than those in the
constant case since the decay starts very early.  This demonstrates
the efficiency of the rule developed here which is applicable to
practical scenarios.
\newline

\underline{7. Conclusion}

A statistical mechanics approach to optimising on-line principal
component analysis provides insight to the learning process. The
theoretically obtained time-dependent optimal learning rates depend on
quantities which are not accessible in practical applications;
however, examining the optimal learning scenarios led to the
development of a practical technique for speeding up the training
process on the basis of observables that can be easily monitored in
practical scenarios. The new method has been demonstrated on a simple
problem and was shown to improve the training performance
considerably.

\vspace{5mm} {\small {\bf Acknowledgments} \quad This work has been
partially support by the EU grant CHRX-CT92-0063 and the British
Council grant: British-German Academic Research Collaboration
Programme project 1037. DS also acknowledges support from the
Leverhulme Trust (F/250/K). ES and DS would like to thank Magnus
Rattray for usefull discussions and suggestions.}


\newpage

\begin{thebibliography}{10}
\bibitem{Hertz} J Hertz, A Krogh and R Palmer {\it Introduction to the
             Theory of Neural Computation} , Addison--Wesley, Redwood
             City, CA (1991)
%
\bibitem{Bishop} C Bishop {\it Neural Networks for Pattern Recognition},
                   Clarendon Press, Oxford (1995)
%
\bibitem{Bi94}  M Biehl, Europhys. Lett. {\bf 25} (1994) 391
                         Neurocomp.\ {\bf 5} (1993) 185
%
\bibitem{Sa98} D Saad (ed) {\it On-line Learning in Neural Networks},
Cambridge University Press, Cambridge UK (1998)
%
\bibitem{WaAl93} T Watkin, A Rau, and M Biehl, Rev.Mod.Phys. {\bf 65}
(1993) 499
%
\bibitem{Sanger} T Sanger, Neural Networks {\bf 2} (1989) 549
%
\bibitem{BiSch98} M Biehl and E Schl{\"o}sser, J. Phys. A {\bf 31}
(1998) 79
%
\bibitem{KiCa92} O Kinouchi and N Caticha, J. Phys. A {\bf 25} (1992)
  6243
%
\bibitem{SaRa97} D Saad and M Rattray, Phys. Rev. Lett. {\bf 79}
  (1997) 2578 ; M Rattray and D Saad, Phys. Rev. E {\bf 58} (1998) 6379
%
\bibitem{RaSa97} M Rattray and D Saad, J. Phys. A {\bf 30} (1997) 771
\end{thebibliography}
\end{document}